\documentclass{aastex}
\usepackage{graphicx}
\usepackage{epstopdf}
\usepackage{lscape}
\usepackage{natbib}  
\usepackage{hyperref}
\usepackage[caption=false]{subfig}
\usepackage{spr-astr-addons}
\usepackage{url}
\usepackage{enumitem}
\usepackage{float}
\usepackage{tikz}
\usetikzlibrary{shapes.geometric, arrows}
\usepackage{algorithm,algorithmic}

\tikzstyle{startstop} = [rectangle, rounded corners, minimum width=3cm, minimum height=0.5cm,text centered, draw=black, node distance=2.5cm,  fill=white!30]
\tikzstyle{io} = [trapezium, trapezium left angle=70, trapezium right angle=110, minimum width=3cm, text width=5cm, minimum height=0.5cm, text centered, draw=black, node distance=2.5cm, fill=white!30]
\tikzstyle{process} = [rectangle, minimum width=5cm, minimum height=1cm, text centered, text width=7cm, draw=black, node distance=2.5cm, fill=white!30]
\tikzstyle{decision} = [diamond, minimum width=0.5cm, minimum height=0.5cm, text centered, text width=1.5cm, draw=black , node distance=5.5cm, fill=white!30]
\tikzstyle{arrow} = [thick,->,>=stealth]

\RequirePackage{color}
\begin{document}

\title{Spectral Clustering for Optical Confirmation and Redshift Estimation of  X-ray Selected Galaxy Cluster Candidates in the SDSS Stripe 82}


\author{Eman Mahmoud\altaffilmark{1,2}}
\email{eman.ahmed@ejust.edu.eg} 

\author{Ali Takey\altaffilmark{2,3}}

\and

\author{Amin Shoukry\altaffilmark{1}}

\altaffiltext{1}{Egypt-Japan University of Science and Technology (E-JUST), 21934 New Borg El-Arab, Alexandria, Egypt.}
\altaffiltext{2}{National Research Institute of Astronomy and Geophysics (NRIAG), 11421 Helwan, Cairo, Egypt.}
\altaffiltext{3}{Sorbonne Universit\'es, UPMC Univ.~Paris 6 et CNRS, UMR~7095, Institut
d'Astrophysique de Paris, 98bis Bd Arago, 75014 Paris, France}



\begin{abstract}
We develop a galaxy cluster finding algorithm based on spectral clustering technique to identify optical counterparts and estimate optical redshifts for X-ray selected cluster candidates\footnote{\url{https://github.com/1680/Journal-manuscript-code.git}}.  
As an application, we run our algorithm on a sample of X-ray cluster candidates selected from the third XMM-Newton serendipitous source catalog (3XMM-DR5) that are located in the Stripe 82 of the Sloan Digital Sky Survey (SDSS). Our method works on galaxies described in the color-magnitude feature space. We begin by examining 45 galaxy clusters with published spectroscopic redshifts in the range of 0.1 to 0.8 with a median of 0.36. As a result, we are able to identify their optical counterparts and estimate their  photometric redshifts, which have a typical accuracy of 0.025 and agree with the published ones. 
Then, we investigate another 40 X-ray cluster candidates (from the same cluster survey) with no  redshift information in the literature and found that 12 candidates are considered as galaxy clusters in the redshift range from 0.29 to 0.76 with a median of 0.57. These systems are newly discovered clusters in X-rays and optical data. Among them 7 clusters have sepectroscopic redshifts for at least one member galaxy. 
\end{abstract}




\section{Introduction}
\label{Sec:Intro}

Study of galaxy clusters is an important field in Astronomy since they attain maximum values in the density space and their properties  provide one of the ways to test expansion and structure growth models \citep[e.g.][]{Voit05, Allen11}. Therefore, various cluster surveys have been conducted at X-rays, optical and mm wavelengths due to multi-components of their baryonic matters (galaxy and intracluster gas).    
Detecting galaxy clusters in optical data is not an easy task since the available spatial information is precise only in two dimensions which are the right ascension($\alpha$) and declination($\delta$). The third important spatial dimension is along the line of sight but unfortunately it has large uncertainty, which leads to confusion in detecting galaxy clusters and determining their memberships \citep[e.g.][]{Gal2008, Hao10}.

X-ray cluster surveys provide an accurate selection method for detecting galaxy clusters at wide redshift range up to 1.7 \citep[e.g.][]{Piffaretti11, Takey11, Mehrtens12, Clerc12, Takey13, Takey14, Clerc14, Pierre15}.
Although X-ray observations are not available for a large area on the sky and do not provide redshift for the majority of the selected clusters, they determine precisely the cluster centers and provide observable parameters (X-ray luminosity and temperature) correlating well with the cluster mass. Optical and NIR observations provide the main data to estimate cluster redshifts.  
Galaxy clusters are tightly related in space ($\alpha$,~$\delta$ and redshift) as well as in color. Therefore, optical cluster detection algorithms try to make use of these properties to identify galaxy clusters and their members as well as to estimate cluster redshifts \citep[e.g.][]{Koester07, Hao10, Rykoff14, Durret15, Wen15}.

In this paper, we use the spectral clustering algorithm for optically detecting and estimating photometric redshift of X-ray selected cluster candidates from the 3XMM/SDSS Stripe 82 galaxy cluster survey \citep{Takey2016}. 
Our cluster finding algorithm tries to identify galaxy clusters 
in the multi-dimensional color-magnitude space of galaxies. An important goal of the algorithm is to minimize both false positive overheads (projection effects) and false negative detections (missing cluster galaxies) which lead in turn to precisely estimate the cluster redshift.
Applying the algorithm on a subsample of 45 clusters with available spectroscopic redshifts in the literature gives recovery for all of them with comparable photometric redshift estimates. In addition, we optically confirm and estimate redshifts of 12 X-ray cluster candidates which had no available redshifts in literature. Those are considered as new galaxy clusters in the literature.

The paper is organized as follows. Section~\ref{Sec:RelatedWork} gives a brief description of related work in optical detection of galaxy clusters. We briefly describe the 3XMM/SDSS Stripe 82 galaxy cluster survey and its X-ray and optical data in section~\ref{Sec:DataDescr}, while the cluster samples studied in this work are presented in section~\ref{Sec:DataInvest}.  
Section~\ref{Sec:Method} presents the spectral clustering-based galaxy clusters finding algorithm, which we develop for photometric redshift estimation and cluster confirmation. Section~\ref{Sec:Results} gives the results for the  studied cluster samples and a comparison with the published ones. 
We end by section~\ref{Sec:Dis} for conclusion and future work.
Throughout this paper we use the following values for the cosmological constants: $H_0$=70 $km~s^{-1} Mpc^{-1}$, $\Omega_M$=0.3, and $\Omega_\lambda$=0.7.



\section{Related work in optical detection of galaxy clusters}
\label{Sec:RelatedWork}

\citet{Gal2008} gives an overview of optical detection algorithms of galaxy clusters. These techniques are developed to detect galaxies clustered in three spatial dimensions after de-projection of field galaxies. Therefore, these algorithms can be classified  based on their de-projection method using available data as summarized by \citet{Hao10}. At first, only single-band data were available, but with advance in the digital imaging technology, multi-band data became available as well.

For a single-band data, the clustering algorithms depend only on the magnitude property of the galaxies as in smoothing kernels \citep{Shectman85}, adaptive kernel \citep{Gal00}, matched filter \citep{Postman96}, voronoi tessellation \citep{Kim02} and cut and enhance algorithm \citep{Goto02}.
The single-band magnitude of galaxies could be used for detection of massive clusters.
The constructed cluster catalogs based on single-band data do not have good purity nor completeness for low/intermediate mass regimes \citep{Hao10}.   

With the availability of multi-band data, clustering algorithms
began to use either colors or photometric redshifts of galaxies. Cluster galaxies can be identified using the red sequence (E/S0 ridgeline), which has a very narrow color scatter and slightly tilted color-magnitude relation \citep[e.g.][]{Gladders05, Koester07, Hao10, Rykoff14}. 
The other method to de-project galaxies along the line of sight is based on the photometric redshifts of galaxies, which are determined based on their magnitudes or colors. The accuracy of cluster photometric redshifts in this method is about 0.02-0.03 \citep[e.g.][]{Wen09, Takey13, Wen15}. Cluster algorithms based on spectroscopic redshifts of galaxies provide more accurate redshifts for clusters \citep[e.g.][]{Miller05, Berlind06}, but are limited to clusters with available spectroscopic coverage.

In this work, we make use of the multi-band magnitudes available in SDSS as well as the photometric redshifts of galaxies. First, we use our spectral clustering-based cluster finding algorithm to identify galaxies with similar colors including the brightest galaxy within a certain angular radius from the X-ray positions. Then, we use their photometric redshifts to determine the cluster redshift. The available spectroscopic redshifts of galaxies are used to compute the uncertainties of the cluster photometric redshift estimates.   





\section{Data Description}
\label{Sec:DataDescr}

Selecting galaxy clusters in X-rays provides complete and pure cluster samples. X-ray observations provide information about cluster positions and fluxes. Unfortunately, it does not provide redshift information of clusters except for very X-ray bright clusters (in this case, redshift can be measured from X-ray data alone). In our work, we begin with analyzing X-ray (XMM-Newton) observations and then using the available optical information from SDSS Stripe 82 photometric data. In this section, we briefly describe the X-ray cluster candidates as well as the optical data used in our analysis.         

\subsection{X-ray cluster candidates in Stripe 82 region}

We have conducted a galaxy cluster survey based on XMM-Newton observations that are located in the SDSS Stripe 82 footprint \citep{Takey2016}. Due to the small number of XMM-Newton observations pointed at the Stripe 82 region (S82, hereafter), the survey area is 11.25 deg$^2$ corresponding to 74 observations considered in our survey. The X-ray cluster candidates were selected from extended sources in the third XMM-Newton serendipitous source catalog\footnote{\url{http://xmmssc.irap.omp.eu/Catalogue/3XMM-DR5/3XMM_DR5.html}} \citep[3XMM-DR5, ][]{Rosen15}. After visual inspection of both X-ray and optical images of the selected candidates, the X-ray cluster candidate list comprises 94 objects. We have searched for available redshifts in the literature for these candidates. As a result, 54 candidates are previously known as galaxy clusters with available redshifts in the range of 0.05-1.2. The remaining 40 systems are cluster candidates that have no information in the literature about being galaxy clusters and there is no redshift information available. The majority of these candidates are expected to be at high redshifts $ z>0.6$, as indicated from the appearance of surrounding galaxies for X-ray positions in optical images.    

\subsection{SDSS Stripe 82 optical  data}

\citet{Annis14} constructed and characterized the coaddition SDSS 
S82 $ugriz$ imaging data based on repeated scans by SDSS. The data covers 270 deg$^2$ on the sky ($-50.0 \degr  \le \alpha \le 60.0 \degr $ and $-1.25 \degr  \le \delta \le 1.25 \degr $) 
 and reaches 2 magnitude deeper than the normal SDSS single scan data. The galaxy catalog ($\sim$~13 million galaxies) extracted from the co-added deep data is 50$\%$ completeness at $r \sim 23.5$ and $i \sim 23.0$ mags. The photometry is good to 1$\%$ in $u$ and $z$ bands and to 0.5$\%$ in $g$, $r$, and $i$ bands. The data was made public in the Catalog Archive Server (CAS) of SDSS-DR7 and included in the successive releases.
   
\citet{Reis12} have estimated the photometric redshifts of galaxies ($\sim$~13 million objects) that are detected in the co-add data in S82 region by \citet{Annis14}. They used the artificial neural network technique to estimate the photometric redshifts of galaxies in $16 < r < 24.5$ and the nearest neighbor error method to determine the errors in the estimated photometric redshifts. Comparing the estimated photometric redshifts to the training galaxy sample with spectroscopic redshifts indicated that the photometric redshift error is smaller than 0.031.~The catalog is available as an SDSS value-added product of SDSS-DR7\footnote{\url{http://classic.sdss.org/dr7/products/value_added/index.html}}.

The recent data release of the SDSS projects is the SDSS-DR12\footnote{\url{http://skyserver.sdss.org/dr12/en/tools/chart/navi.aspx}} \citep{Alam15}. It includes all SDSS observations through July 2014. One of the important catalogs among the released data is the galaxy spectra catalog that comprises $\sim$~2.4 million galaxies with spectroscopic redshifts. These spectra were performed in the framework of the Baryon Oscillation Spectroscopic Survey (BOSS) and previous SDSS spectroscopic programs. The galaxy spectroscopic redshifts are available in the CAS of SDSS-DR12.

In our present work, we use the photometric data from the S82 co-add data \citep{Annis14}, photometric redshifts of galaxies and their errors from \citep{Reis12}, and the spectroscopic redshift of galaxies in the SDSS-DR12 \citep{Alam15}. First, we obtained the photometric parameters (IDs, positions, dereddened model magnitudes in $griz$ bands and their errors) of galaxies surrounding X-ray cluster candidates from S82 co-add data ({\tt Galaxy} view table in CasJobs). In this step, we used a similar SDSS query and a magnitude cut ($16 < r < 24.5$) as the ones used by \citet{Reis12}. Second, we cross correlated the extracted galaxy samples with the galaxy photometric redshifts (Reis catalog) through the IDs of galaxies since they are the same in both catalogs. Third, the spectroscopic redshifts, when available, were retrieved from SDSS-DR12 ({\tt SpecObj} table in CasJobs) through cross-matching the positions of galaxies since the IDs in SDSS-DR7 and -DR12 are different. 

Finally, for each X-ray cluster candidate in our list (94 candidates) we have a galaxy table including all galaxies within 10 arcmin ($\sim$~1100 kpc at a redshift of 0.1) from its center with $IDs$, $\alpha$, $\delta$, $griz$ mags and their errors, $z_{phot}$, $err\_z_{phot}$ (Reis catalog), and $z_{spec}$ (SDSS-DR12), when available. We also created galaxy tables in the same way, as mentioned above, for 100 random positions in S82 region to estimate the galaxy density of fore- and background galaxies in a physical radius as a function of redshift to be used for testing and comparing our results.          

The galaxy sample is in the magnitude range of $16<r<24.5$.  
To filter the extracted galaxies from very faint objects, we excluded all galaxies fainter than 25 mags in the other bands ($giz$). 
We also remove galaxies with large relative errors in all magnitudes higher than 0.1. In the code of our algorithm, we  apply another magnitude cut of $i < 23$ (50$\%$ completeness magnitude as given by \citet{Annis14}) with an error smaller than 0.3.   

\citet{Reis12} recommended to apply a cut on the error of $z_{phot}$ for high redshift galaxies ($>0.75$) since faint galaxies have high redshift errors. We used galaxies with redshift $0.03<z_{phot}<1$ and relative error $(err\_z_{phot}/z_{phot} )$ smaller than 0.5. The objects in the galaxy sample that are classified as 'Star' from their spectra (as given in {\tt SpecObj} table in SDSS-DR12) are also excluded.  



\section{Study Sample Investigation}
\label{Sec:DataInvest}

\begin{figure*}[t]
\centering
\includegraphics[scale=0.45]{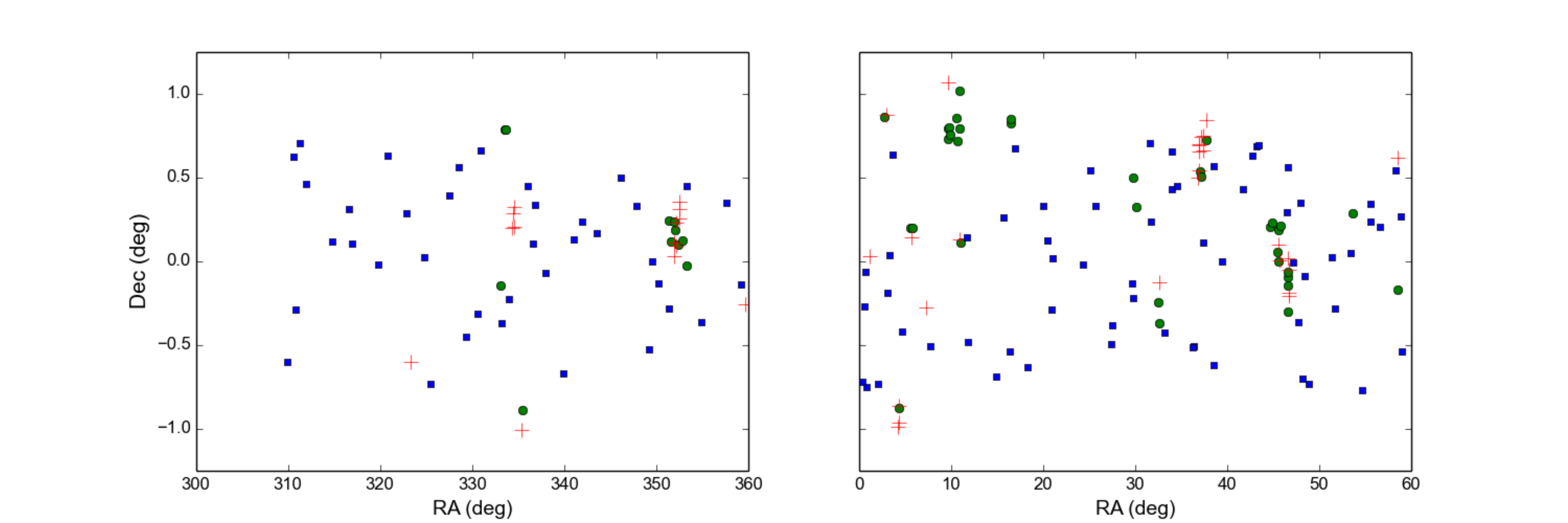}
\caption{Sky distribution of study samples investigated in our work: green dots, red pluses and blue squares correspond to known galaxy clusters (45 systems), unconfirmed cluster candidates (40 objects) and random positions (100 locations), respectively.} 
\label{Fig:Stripe82}
\end{figure*}

\begin{figure*}[t]
\centering
\includegraphics[width=5.7cm]{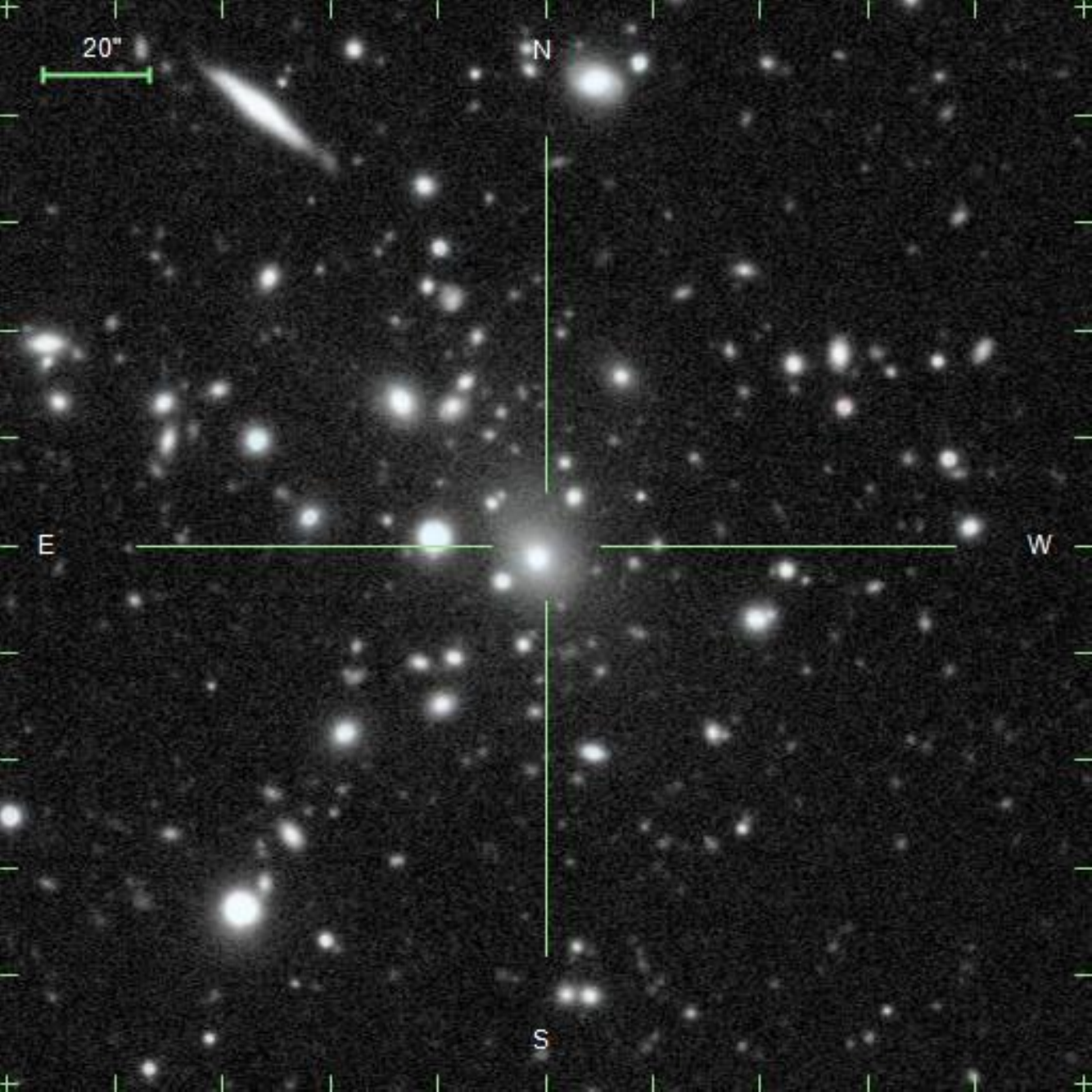}
\includegraphics[width=5.7cm]{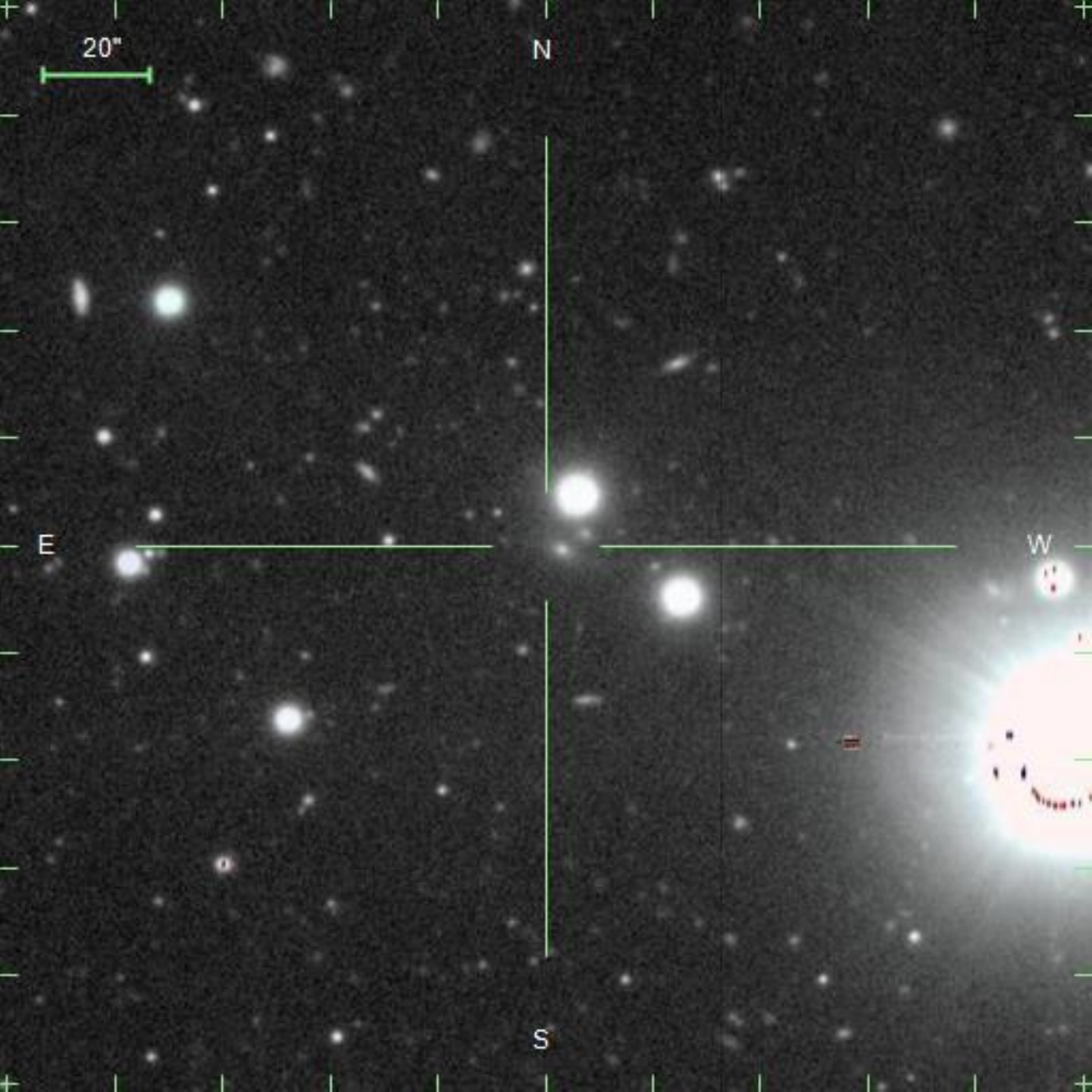}
\includegraphics[width=5.7cm]{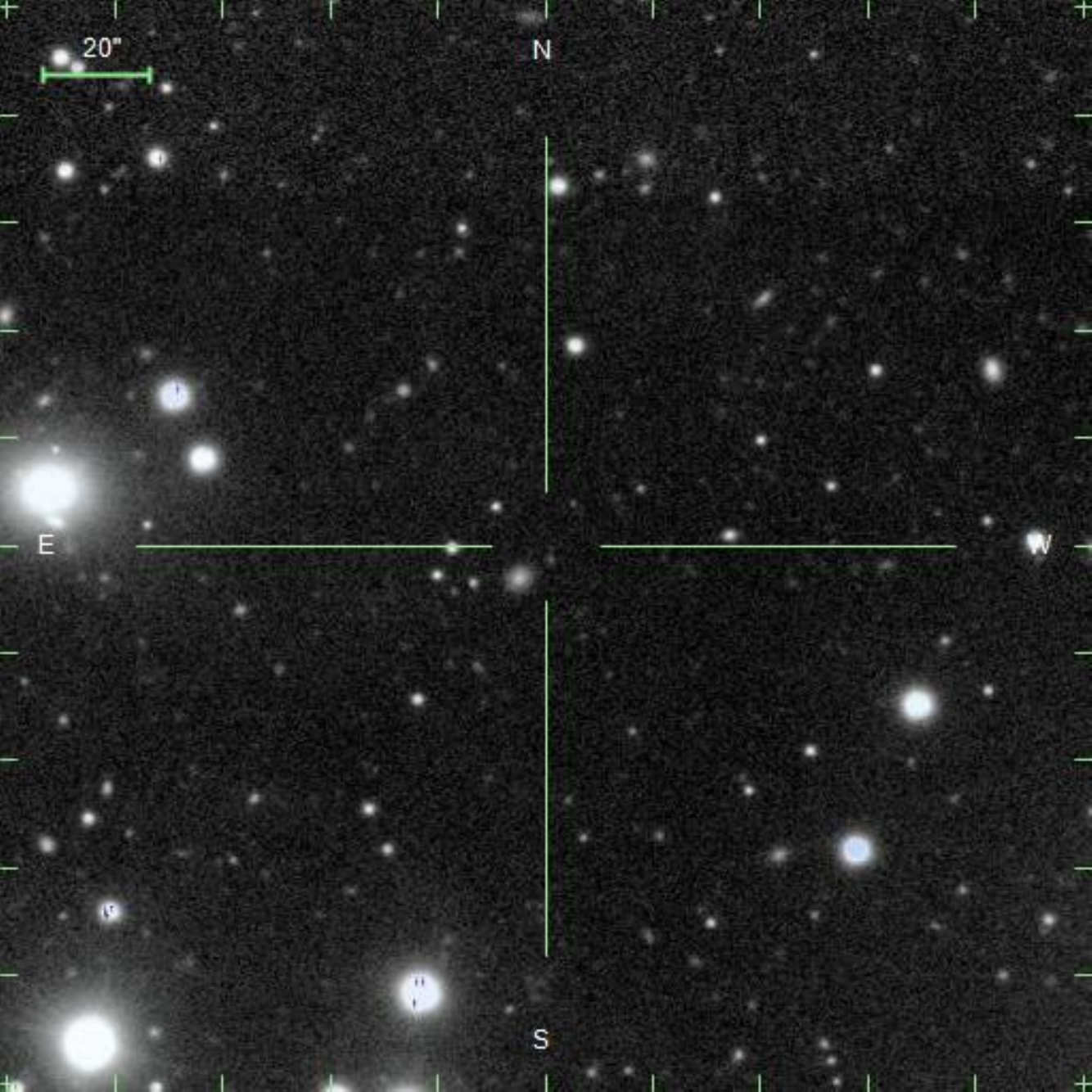}
\caption{SDSS  deep images of examples of a known cluster at $z=0.2141$ (left, 3XMM J001737.3-005240), a distant cluster candidate (middle,  3XMM J030213.0+000559), a random position (right, $\alpha$=327.495172,  $\delta$=0.396132) in S82 region. Left and middle images are centered at X-ray emission peaks (marked by the cross-hairs) while the right one is centered at one position of the 100 random positions, see the text.}
\label{Fig:ClusterExample}
\end{figure*}

We aim to optically confirm and to estimate redshifts of the X-ray cluster candidates (94 systems) using our cluster detection algorithm based on the available photometric data from the SDSS S82 survey. As mentioned above, of these candidates, 54 systems are previously known as galaxy clusters in the literature with available redshifts in the range of 0.05-1.2. These available redshifts are important to validate the redshifts estimated from our work. 

Due to the magnitude limits of S82 and our magnitude cut ($i<23$), we are unable to detect galaxy clusters beyond redshift of 0.8. Also, very low redshift clusters subtend large angular area on the sky that are not suitable for our algorithm, which investigates nearby cluster galaxies to the X-ray position. Therefore, we have excluded two high redshift clusters at $z>0.8$ and three low redshift systems at $z<0.1$. Additionally, we also excluded 4 systems that have only photometric redshifts or not completely covered in S82 survey. We are left with a cluster sample comprising 45 clusters with spectroscopic redshifts in the range of 0.1-0.8. This sample is used in our algorithm to investigate and validate our results.

After having acceptable redshifts for the known clusters, we go ahead applying the algorithm on the remaining X-ray cluster candidates (40 systems) that have no redshift information in the literature. We do not expect to identify those candidates that are more probable to be at high redshifts, i.e. at $z > 0.8$. To assess our optical cluster detections of the X-ray cluster candidates, we generated a list of 100 random positions in S82 region and checked their clustering galaxies.

Fig.~\ref{Fig:Stripe82} shows the sky distribution of our study samples: known galaxy clusters with spectroscopic redshifts (45 systems), X-ray cluster candidates (40 objects) with no available redshifts, and our chosen 100 random positions. Fig.~\ref{Fig:ClusterExample} shows an example of confirmed cluster sample (3XMM J001737.3-005240 at $z=0.2141$), cluster candidate sample (3XMM J030213.0+000559), and random positions ($\alpha$=327.495172, $\delta$=0.396132). 
We can clearly see an overdensity of galaxies around the X-ray  center of the low redshift cluster (left) and low galaxy density around the distant cluster candidate (middle) and the random position (right).



\section{The Method}
\label{Sec:Method}

\subsection{Spectral Clustering Technique}
\label{SubSec:MethodSpectralClustering}

Spectral clustering algorithm is a general technique that partitions the rows of an affinity matrix according to their components in the top k singular vectors of the matrix \citep{Kannan2004}.~If we consider the rows of the matrix as points in a high dimensional space, the spectral algorithm projects all points onto the subspace defined by the top k singular vectors of the matrix.~Each singular vector defines a cluster and each projected point is back-mapped to the cluster defined by the closest angle singular vector.   

In general, clustering techniques are not only limited to computer science applications, but they are also widely applied in different fields of science such as Statistics, Biology, Psychology and Astronomy.~Compared to other clustering techniques, spectral clustering is easily implemented and efficiently solved by standard linear algebra software \citep{Lux2007} which makes it a good choice for many problems. The spectral clustering algorithm is able to identify clusters by the connectivity of its members as well as the density \citep{Lux2007, Kannan2004}.

When applying the spectral clustering algorithm to the galaxy clusters detection problem, galaxies are considered as observations (points in multi-dimensional space) and their properties are considered as features (dimensions of the space). Related galaxies in color and magnitude properties are grouped in the same cluster where their internal similarities are maximized while external similarities with galaxies in other clusters are minimized.

The pairwise similarity between galaxies uses the fully connected similarity measure which is symmetric and non negative and pairs of galaxies are connected based on their colors and magnitudes.~The Gaussian similarity is an example of the pairwise similarity where
\begin{equation} 
 S(G_i,G_j)=exp(-||G_i-G_j||/2 \sigma^2); 
\end{equation}
where $S(G_i, G_j)$ is the similarity measure between $galaxy~G_i$ and $galaxy~G_j$,~$i, j \in [1 ... n]$ and $i \neq j$, if $i = j$, then $S(G_i, G_j)$ is set equal to 0. The number of galaxies is $n$ and $\sigma$ is a parameter that controls the width of the galaxy neighborhood.

In our galaxy cluster finding algorithm we are using the basic spectral clustering\footnote{\url{http://www.mathworks.com/matlabcentral/fileexchange/34412-fast-and-efficient-spectral-clustering}} according to \citep{Ng2002}, setting the number of clusters to $k=2$. The  distance measure in the k sub-dimensional space is adopted by experimentation to be the  $"cosine"$ distance.
The matalb code of our algorithm is available at {\url{https://github.com/1680/Journal-manuscript-code.git}}


\subsection{The Algorithm}
\label{SubSec:MethodAlgor}

Here we present the steps of our spectral clustering-based galaxy cluster finding algorithm. First, applying the spectral clustering on the galaxies within 1 arcmin from the X-ray position to identify  cluster galaxy members. Second, estimating a tentative cluster redshift based on the photometric redshifts of the selected members.
Third, fine-tuning the tentative redshifts based on photometric redshifts of cluster galaxies within 500 kpc from the X-ray center.      
The details of the cluster finding algorithm and its parameters are presented in the pseudo code~\ref{Alg:MyAlg} (Algorithm 1).  The algorithm flowchart in the Appendix (A) shows a summarized visual steps of our cluster finding algorithm. 
In the following subsections, we discuss our adaptation for the cluster spectral algorithm to solve the galaxy cluster finding problem. 


\begin{algorithm*}[t]
\caption{: Spectral clustering-based galaxy cluster finding algorithm}
\begin{algorithmic}[1]
\small
  \STATE For each X-ray cluster candidate do:
\begin{enumerate}[label=1.\arabic*:]
\item Select galaxies satisfying the following constraints:
	\begin{enumerate}[label=\roman*)]
	\item Angular separation from the candidate center $<$ 1 arcmin.
	\item 0.03 $<z_{phot}<$ 1 and $(err\_z_{phot}/z_{phot} )$ $<$ 0.5.
	\item $i<$ 23 and $err\_i <$ 0.3.
	\end{enumerate}

\item For each selected galaxy from the previous step do:

\begin{enumerate}[label=1.2.\arabic*:]
\item Select magnitudes and colors for each galaxy as features (sec.~\ref{SubSec:MethodChoATT}). 
\item Normalize features to the range [0:1].
\end{enumerate}

\item Calculate similarity values between each pair of galaxies based on color and magnitude features. This produces a similarity matrix of dimension ($n \times n$).

\item Use the spectral clustering algorithm to cluster the selected galaxies based on the similarity matrix into $k=2$ clusters (a target cluster and a 'fore- and background' cluster). Choose the cluster having the brightest galaxy in $r$ band to be our target cluster (explained in sec.~\ref{SubSec:MethodK}).

\item Filter the galaxies in the target cluster based on a cluster membership threshold (explained in sec.~\ref{SubSec:MethodGalFilter}).

\item Construct a histogram of photometric redshift estimates of the galaxies obtained from the previous step with a bin resolution of 0.05. Find the redshift corresponding to the peak of the histogram and consider it as the cluster tentative redshift. 

\item Fine tune the cluster tentative redshift as follows:
\begin{enumerate}[label=1.7.\arabic*:]

\item Estimate the angular radius corresponding to projected separation of 500 kpc at the tentative redshift.

\item Select all galaxies with photometric redshift within the redshift interval of ($tentative~redshift \pm 0.04(1+ tentative~redshift)$) and within the chosen physical radius. Then, compute the weighted mean redshift, $\hat{z_p}$, for the selected cluster galaxy members to be the target cluster photometric redshift (sec.~\ref{SubSec:MethodWmean}).

\item For the selected galaxies from the previous step, find those having available spectroscopic redshift values and compute the  weighted mean redshift to be the target cluster spectroscopic redshift $\hat{z_s}$.

\item If the difference between the tentative redshift and $\hat{z_p}$ is $<$ 0.001, end the algorithm and report results. If not, set the tentative redshift equal to $\hat{z_p}$ and repeat steps from 1.7.1 to 1.7.4.

\end{enumerate}
\end{enumerate}

\label{Alg:MyAlg}
\end{algorithmic}
\end{algorithm*}


\subsubsection{Choosing attributes for spectral clustering}
\label{SubSec:MethodChoATT}

Red-sequence  technique is a powerful method for separating cluster member galaxies from field (fore- and background) ones.~The E/S0 ridgeline in color-magnitude relations show up in different color spaces as a function of redshift due to the shifting of 4000 \AA~across SDSS filters.~The suitable colors for cluster redshift ranges 
$0.0-0.43,~0.43-0.7,~ 0.7-1.0$ are $g-r,~r-i,~i-z$, respectively, as given in \citep[Table 2,][]{Hao10}.~Therefore, one needs to know the expected redshift of the cluster, in advance, to decide which color to use.~An additional color ($r-z$)  was proposed for a wider redshift range ($z<1.4$) since it increases monotonically with redshift \citep{High10}.         

Since we have no prior information about the cluster redshift, we decided to use all available photometric information.~As galaxy features, we use 4 magnitudes ($g,~r,~i,~z$) and all possible combinations of colors based on these magnitudes, 6 colors 
($g-r,~r-i,~i-z,~g-i,~g-z,~r-z$).~This gives a final feature vector of size 10 for each galaxy $G$.~The choice of this attribute combination could be suitable for identifying clusters at different redshifts. 

%

\subsubsection{Setting the number of clusters and choosing the target cluster}
\label{SubSec:MethodK}

We set the number of clusters $k$ in the spectral clustering algorithm to 2. One cluster corresponds to our target cluster with member galaxies candidates, while the other one corresponds to both foreground and background galaxies. Since we are grouping galaxies in clusters based on magnitude and color information, we choose the cluster containing the galaxy with minimum $r$ magnitude as a brightest cluster galaxy (BCG) candidate.

\subsubsection{Galaxy filtering using membership probability}
\label{SubSec:MethodGalFilter}

We apply the spectral clustering on the galaxies satisfying constraints mentioned in step 1.1, Algorithm \ref{Alg:MyAlg}. As an output, a distance vector is  returned from each galaxy $G$ to each of the $2$ centers of the clusters ($n \times 2$ matrix). In order to improve the accuracy of the tentative redshift, we try to remove outlier galaxies by calculating the probability that every galaxy $G$ is a member of cluster $k$. This probability is the inverse of the distance $D_k$ between galaxy $G$ and center $k$ divided by the sum of the inverse distances to all $k$ clusters as follows:
\begin{equation}
Probability(G \in k)=\dfrac{{1}/{D_k}}{\Sigma_{k=1}^{2}{{1}/{D_k}}}.
\end{equation}
Then, we remove from the selected cluster all galaxies with membership probabilities less than a certain threshold. By investigating the probability values, which are very high (usually more than 95$\%$), we choose a threshold of 80\% membership probability. As a result of applying this threshold, a median of 9\% of the initially selected cluster galaxies are being excluded.
    
\subsubsection{Choosing weights suitable for determining cluster weighted mean redshift}
\label{SubSec:MethodWmean}

The cluster photometric and spectroscopic redshifts are computed based on its member galaxies photometric and spectroscopic redshifts, respectively. Since galaxy redshifts differ in the accuracy of their estimates, weights should be associated to each galaxy redshift and then the weighted mean redshifts are calculated.



\section{Results and Discussion}
\label{Sec:Results}

We present here the results after applying our algorithm on the cluster sample (45 clusters) with known redshifts as well as the cluster candidate sample (40 candidates). We also use the data 
for 100 random positions to estimate the detection reliability 
of the resulting clusters from our method.

\subsection{Applying our spectral clustering-based cluster finding algorithm to the cluster sample with redshifts}
\label{SubSec:ResConfirmed}
 
\begin{figure*}[t]
\centering
\includegraphics[width=7cm]{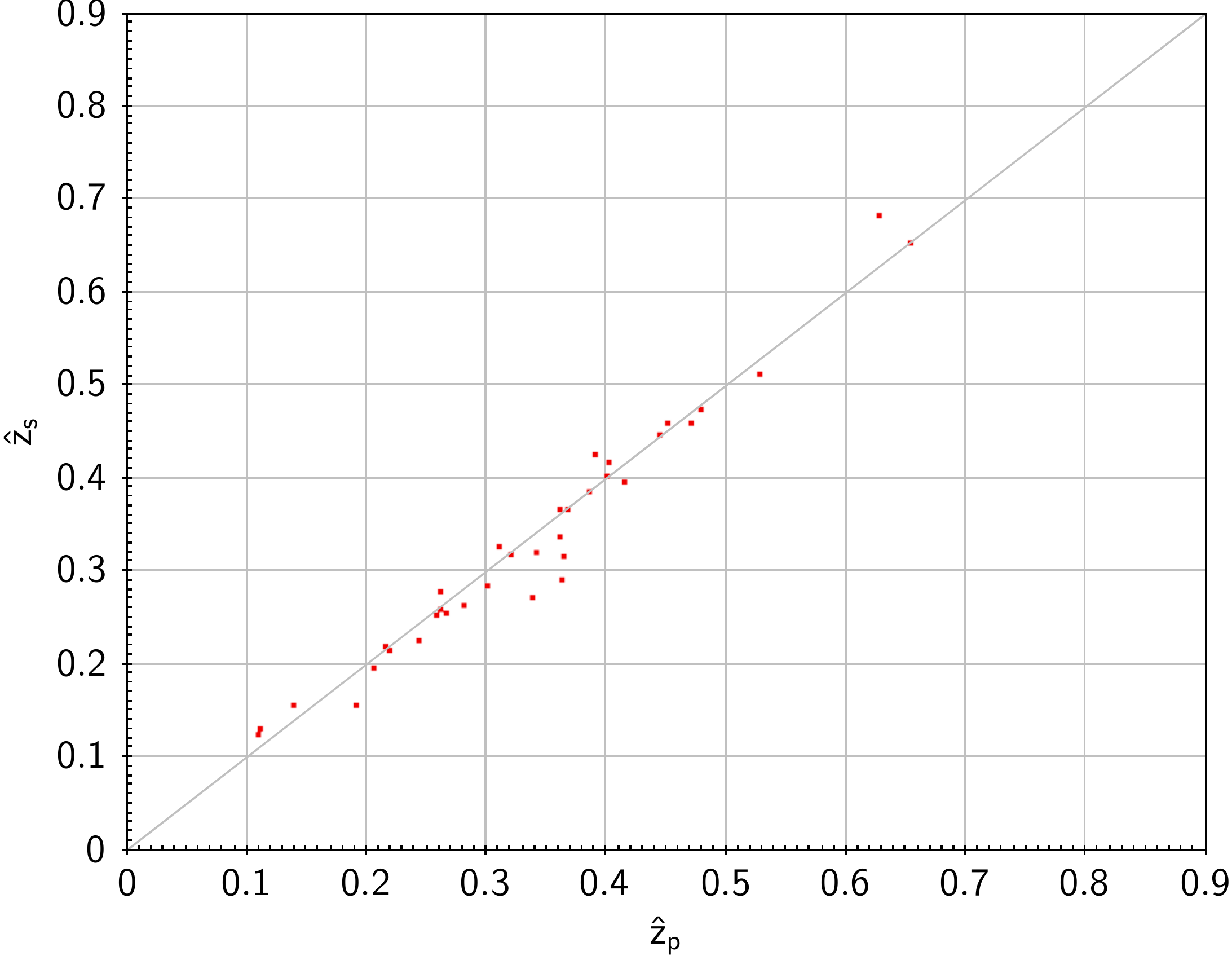}
\hspace{1cm}
\includegraphics[width=7cm]{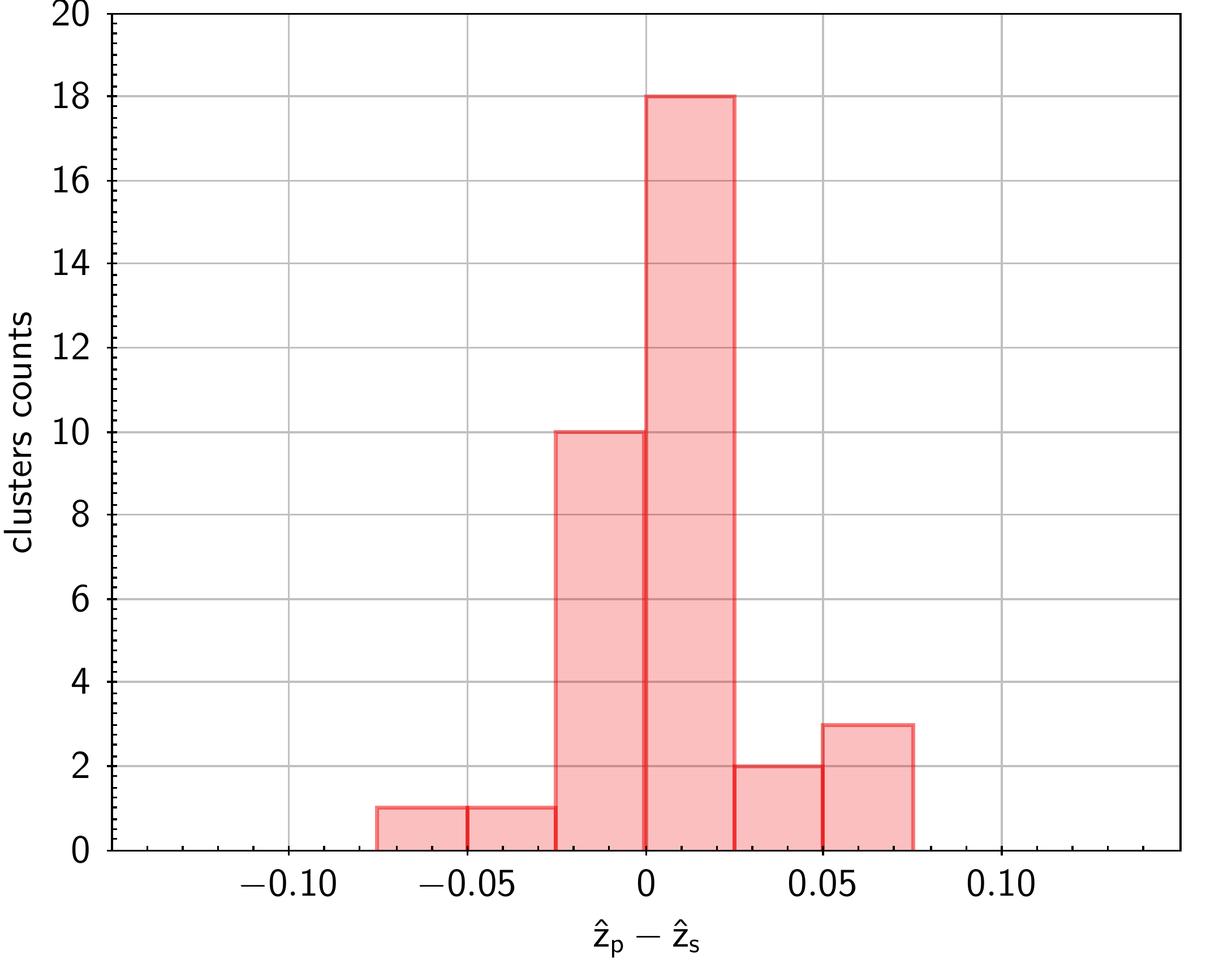}
\caption{Results of the our galaxy cluster finding algorithm: (left) $\hat{z_p}$ versus $\hat{z_s}$ for the cluster sample with known redshifts. Both redshift estimates are derived from our algorithm. The solid line indicates the one to one relationship. (Right) distribution of the differences between $\hat{z_p}$ and $\hat{z_s}$ which have a mean of 0.0078 and a standard deviation of 0.025.}
\label{Fig:zp-zs-comp}
\end{figure*}

\begin{figure*}[t]
\centering
\includegraphics[width=7cm]{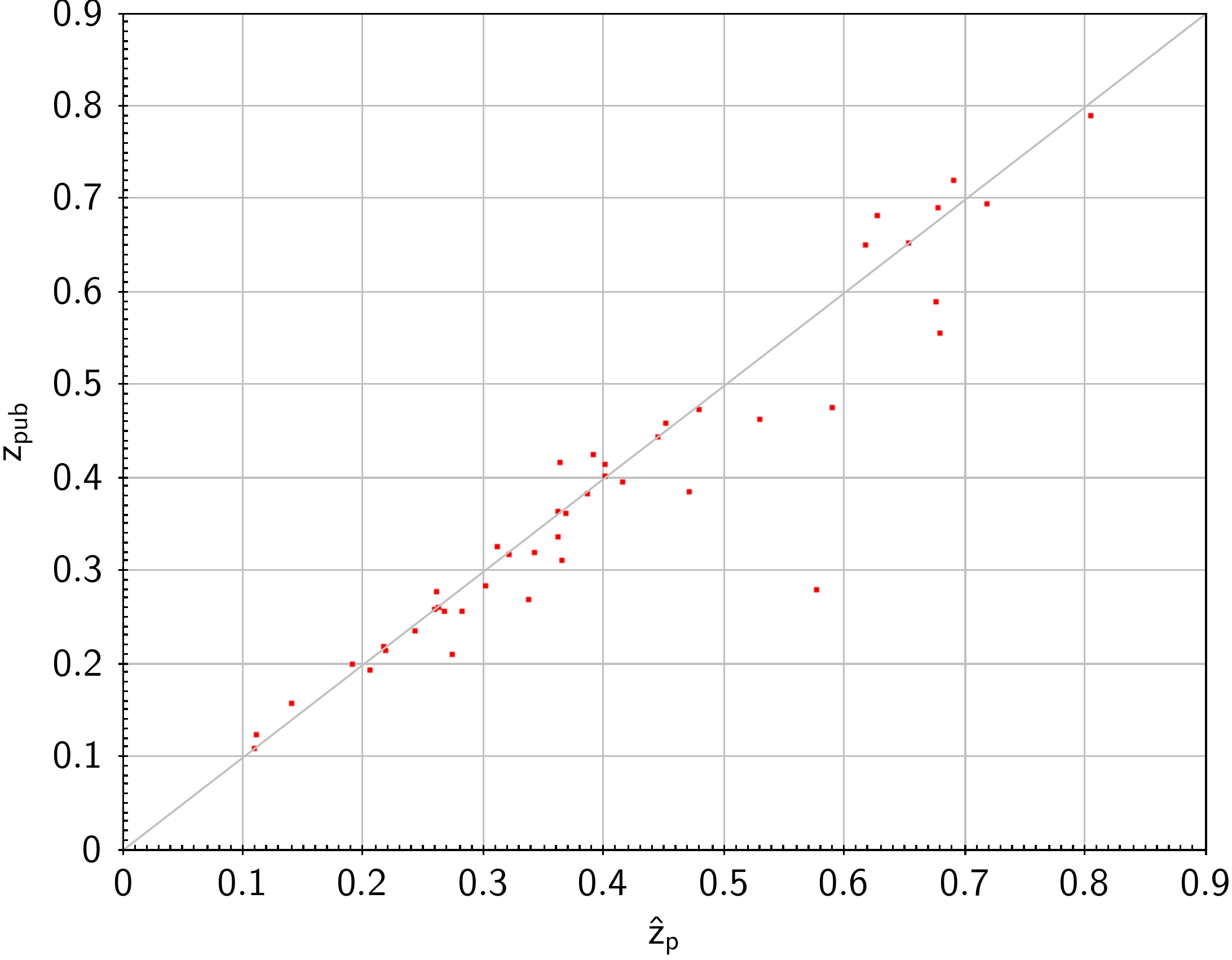}
\hspace{1cm}
\includegraphics[width=7cm]{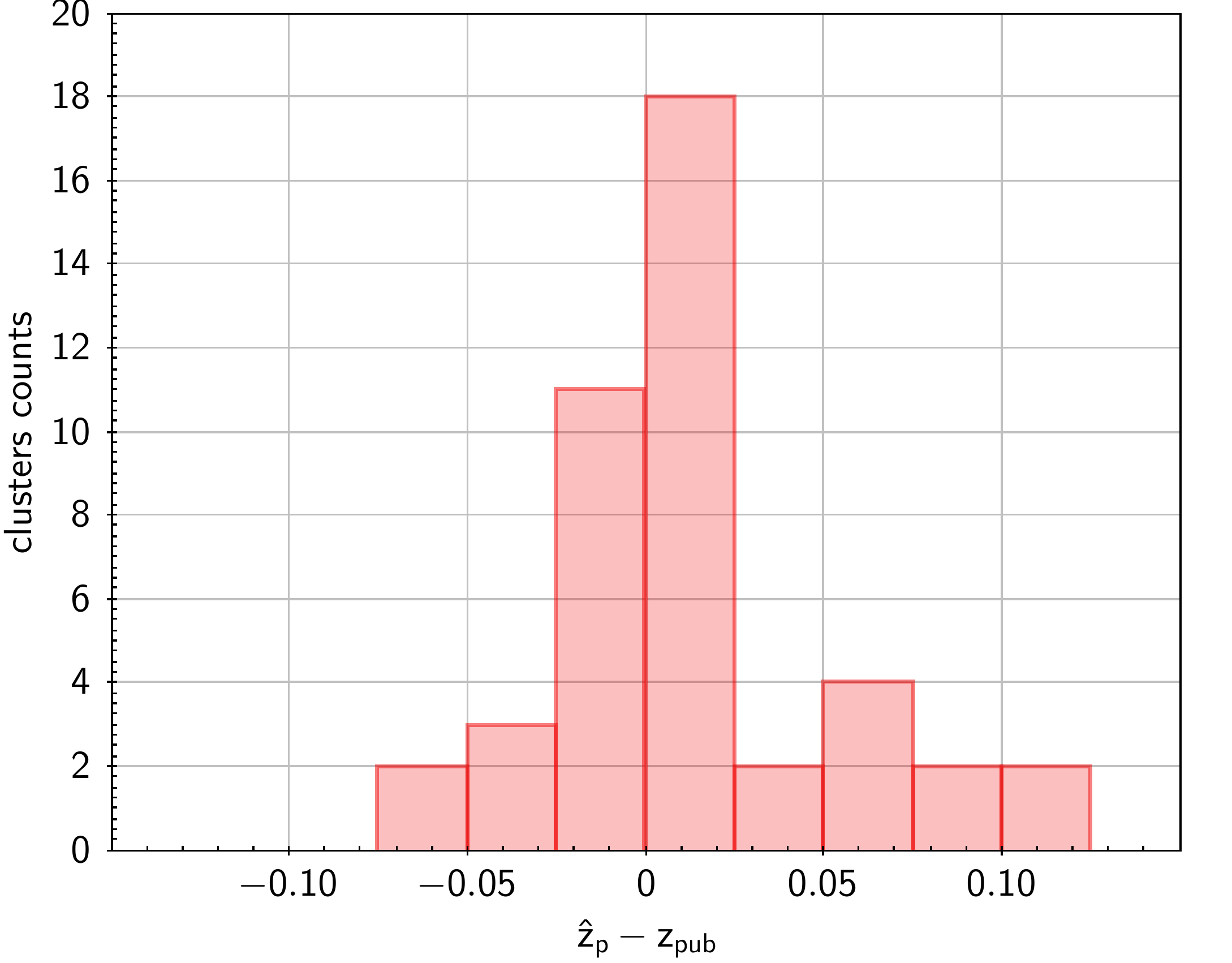}
\caption{Left: $\hat{z_p}$ from our galaxy cluster finding algorithm versus the published spectroscopic redshifts ($z_{pub}$) for the cluster sample. The solid line represents the diagonal of the
square. The differences between the redshift values are shown in the right panel (after excluding the deviated cluster, see text for explanation).}
\label{Fig:ZpZpub}
\end{figure*}

Using our cluster finding algorithm described in sec.~\ref{SubSec:MethodAlgor}, we detect the optical counterparts and estimate photometric redshifts for the cluster sample that are previously known in the literature with spectroscopic redshifts. The photometric redshifts of the cluster sample are re-estimated for the 45 clusters. Of these, 35 systems have spectroscopic confirmation based on at least one member galaxy with spectrum within 500 kpc from the cluster center. Fig.~\ref{Fig:zp-zs-comp} (left) shows the comparison of the photometric redshift estimates ($\hat{z_p}$) and the spectroscopic ones ($\hat{z_s}$) where $\hat{z_p}$ values agree with $\hat{z_s}$ values within $\pm$0.075. The histogram of the differences between these redshifts is shown in Fig.~\ref{Fig:zp-zs-comp} (right). These redshift differences have a mean of 0.0073 and a standard deviation of 0.025. This indicates that the uncertainty in the reported photometric redshift estimates is 0.025.       

We note here that the cluster sample has spectroscopic redshifts either from SDSS or other surveys. When comparing our photometric redshifts with the published spectroscopic ones, we find a good agreement for  $95\%$ accordance within error allowance of $\pm0.1$, as shown in Fig.~\ref{Fig:ZpZpub} (left). The distribution of the redshift differences is shown in Fig.~\ref{Fig:ZpZpub} (right). There is only one cluster (3XMM J002223.3+001201) that has completely different redshift than the published one. By investigating this system, we found a cluster at a redshift of 0.28 identified by \citet[][]{Szabo11, Takey11, Wen15}. Our algorithm detected another cluster at a redshift 0.58 along the line of sight.
After excluding this system, the redshift differences between our estimates and the published ones have a mean of 0.013 and a standard deviation of 0.038.

\begin{figure*}[t]
\centering
\includegraphics[width=7cm]{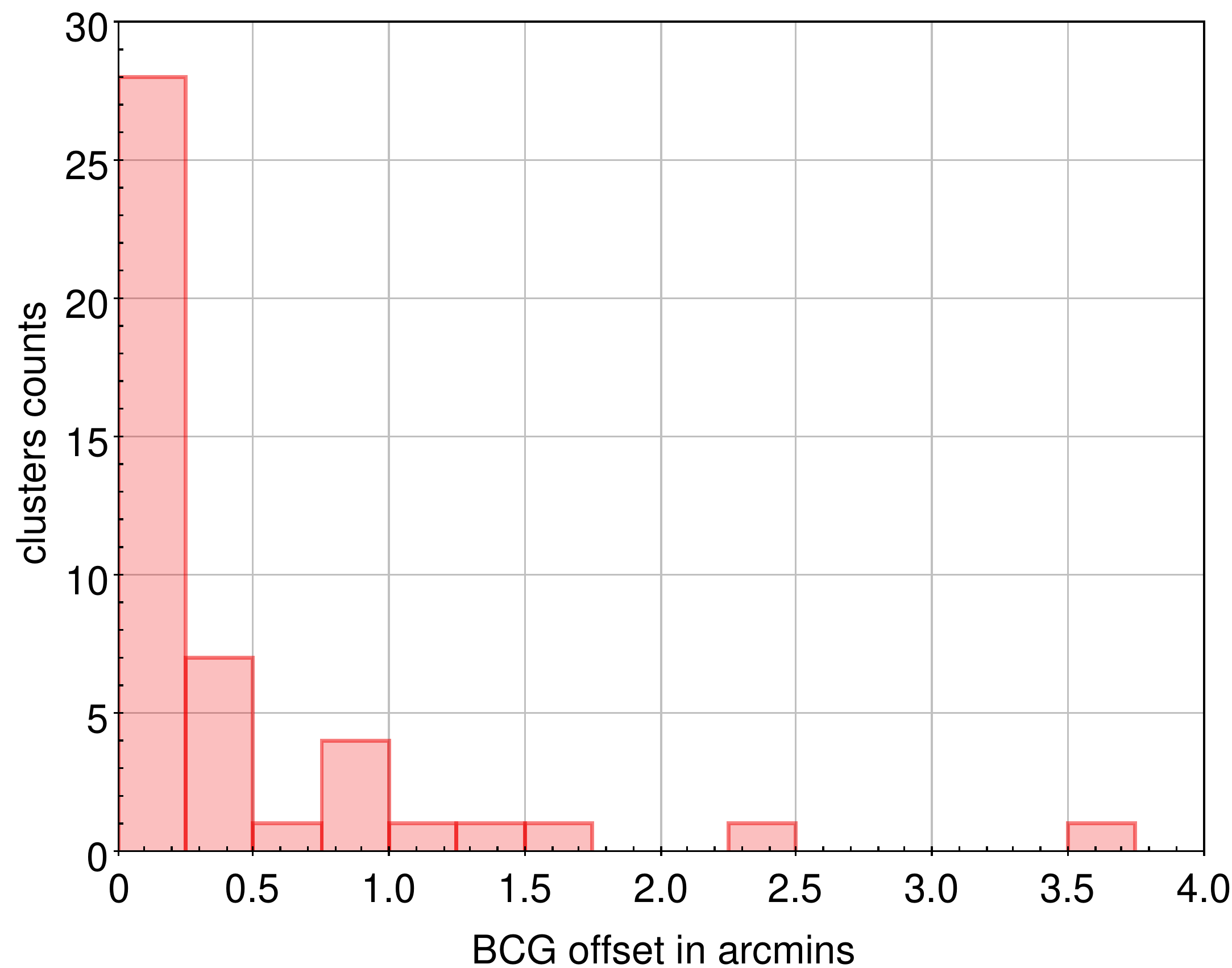}
\hspace{1cm}
\includegraphics[width=7cm]{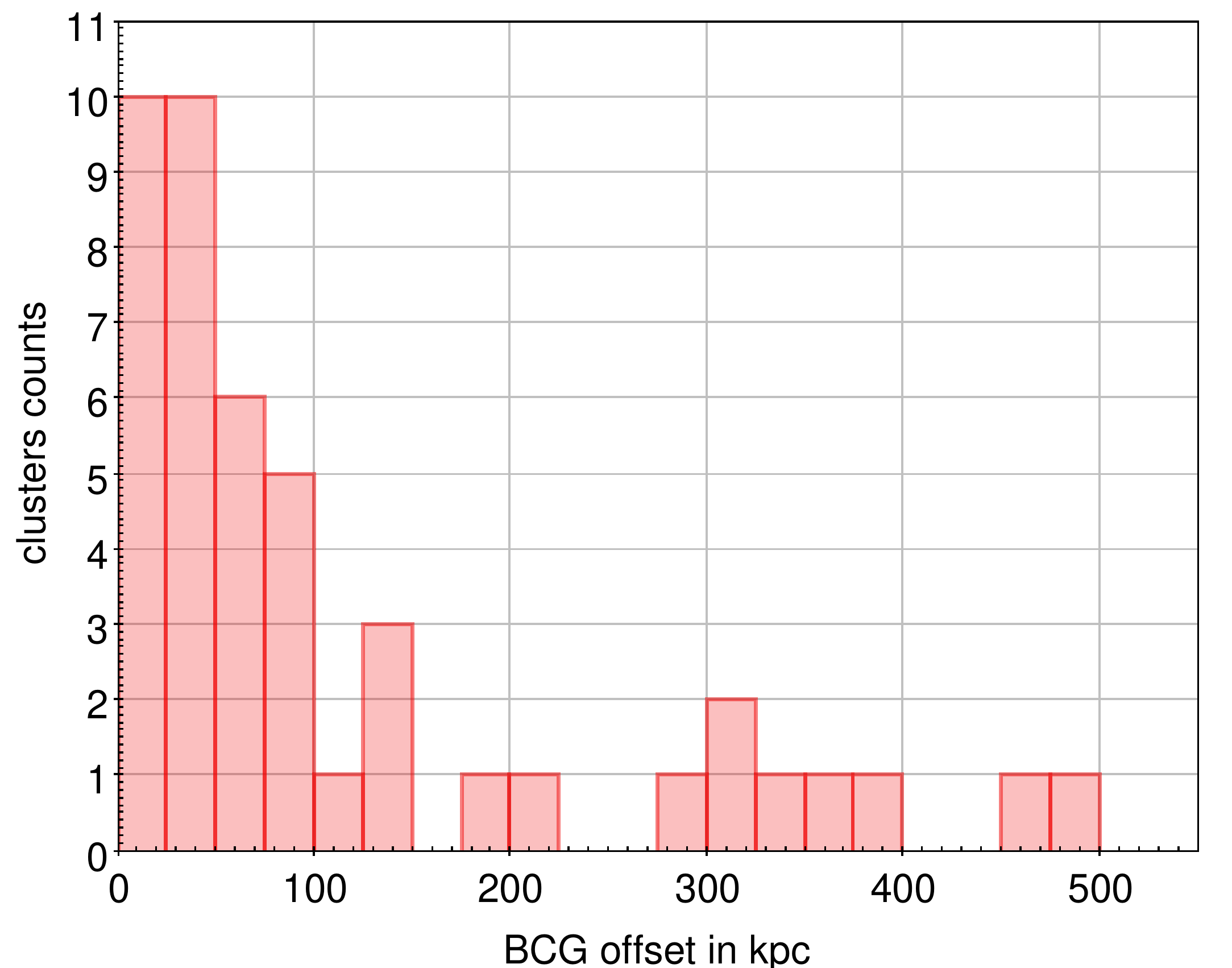}
\caption{Histogram of BCG offset from the X-ray position of the cluster sample in arcmin (left) and in kpc (right). }
\label{Fig:BCGoffset}
\end{figure*}

As mentioned in the algorithm steps, we have selected the BCG as the galaxy with minimum $r$ magnitude within 500 kpc from the X-ray center. The BCG offset is calculated as the separation between its position and the X-ray emission peak position. Fig.~\ref{Fig:BCGoffset} shows the distribution of the BCGs offsets in angular separation (left figure) and physical distance (right one). Cluster with large offset of its BCG might have an ongoing merger.   

Table~\ref{Table:Sample} summarizes our galaxy cluster finding algorithm results for 45 clusters with published spectroscopic redshifts. The columns are: IAUNAME, the International Astronomical Union ID. ($\alpha_{Xray},~\delta_{Xray}$), X-ray detection sky coordinates (J2000, deg). Detection ID, DETID, the X-ray detection number in the 3XMM-DR5 catalog. ($\alpha_{BCG},~\delta_{BCG}$), the BCG coordinates selected within 500 kpc from the X-ray center. 
$n_{z_p}$ is the number of galaxies with photometric redshifts within the redshift interval $\pm$0.04(1+ tentative redshift) and within the physical radius. $\hat{z_p}$ is the weighted average photometric redshift of the cluster calculated from the available $n_{z_p}$ galaxies. $n_{z_s}$ is the number of galaxies with spectroscopic redshifts among $n_{z_p}$ galaxies. $\hat{z_s}$ is the weighted average spectroscopic redshift of the cluster calculated from the available $n_{z_s}$ galaxies. $z_{pub}$ is the spectroscopic redshift puplished in literature \citep{Takey2016}.

\subsection{Investigating random positions}
\label{SubSec:ResRandom}

To avoid identifying non clusters as clusters (false positives) by our algorithm, we define the galaxy density (in $arcmin^{-2}$) as the number of cluster galaxies within certain separation from the X-ray center divided by the area within this radius. We compute galaxy densities within 500 kpc and 1 arcmin in the cluster sample to compare them to the ones computed in the 100 random positions. We define $\mu_{500kpc}$, $\sigma_{500kpc}$, $\mu_{1arcmin}$ and $\sigma_{1arcmin}$ as the mean and standard deviations of galaxy densities within 500 kpc and 1 arcmin, respectively and in order.

\begin{figure*}[t]
\centering
\includegraphics[width=8cm, viewport=100 240 500 540, clip]{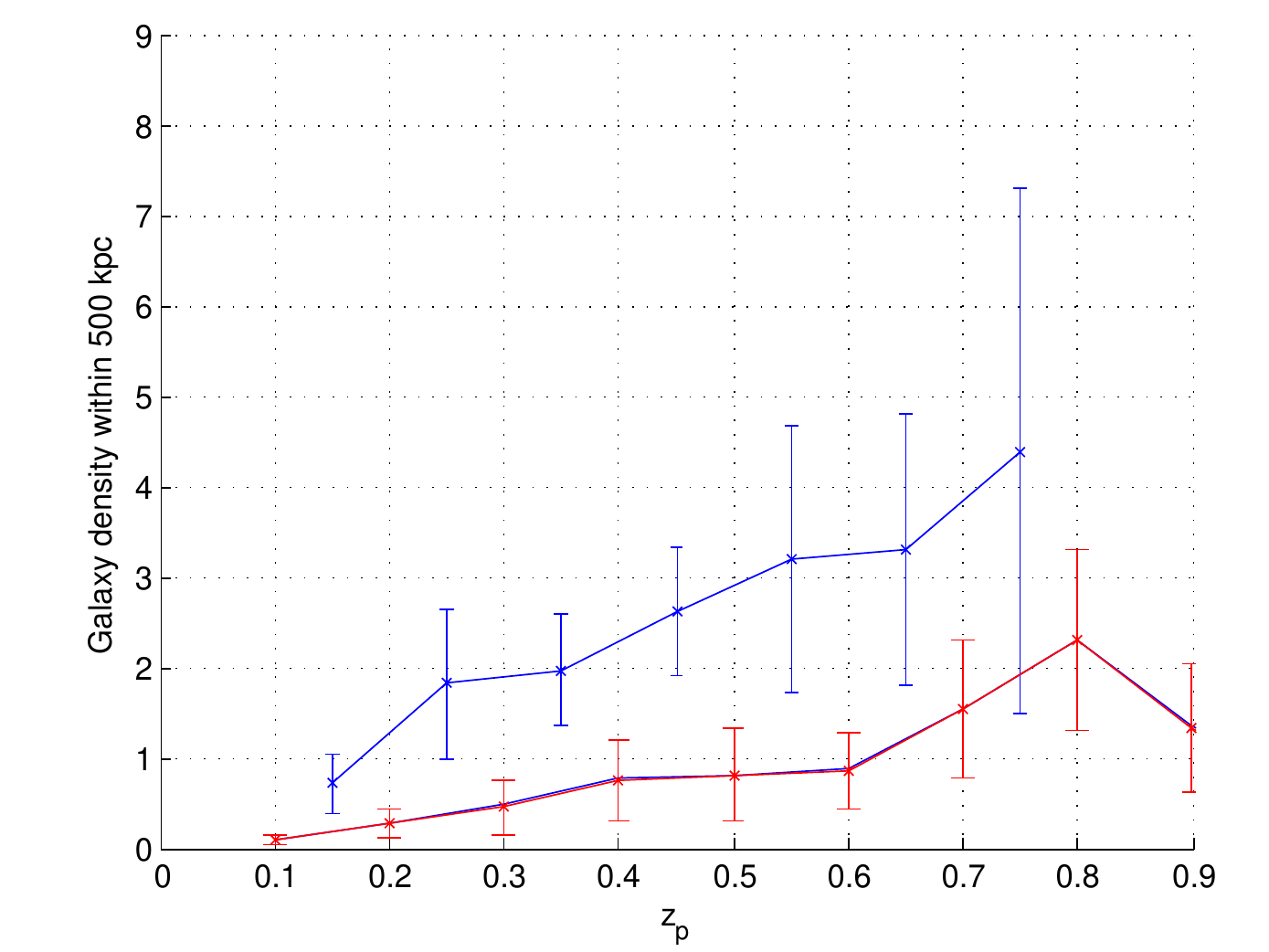}
\hspace{1cm}
\includegraphics[width=8cm, viewport=100 240 500 540, clip]{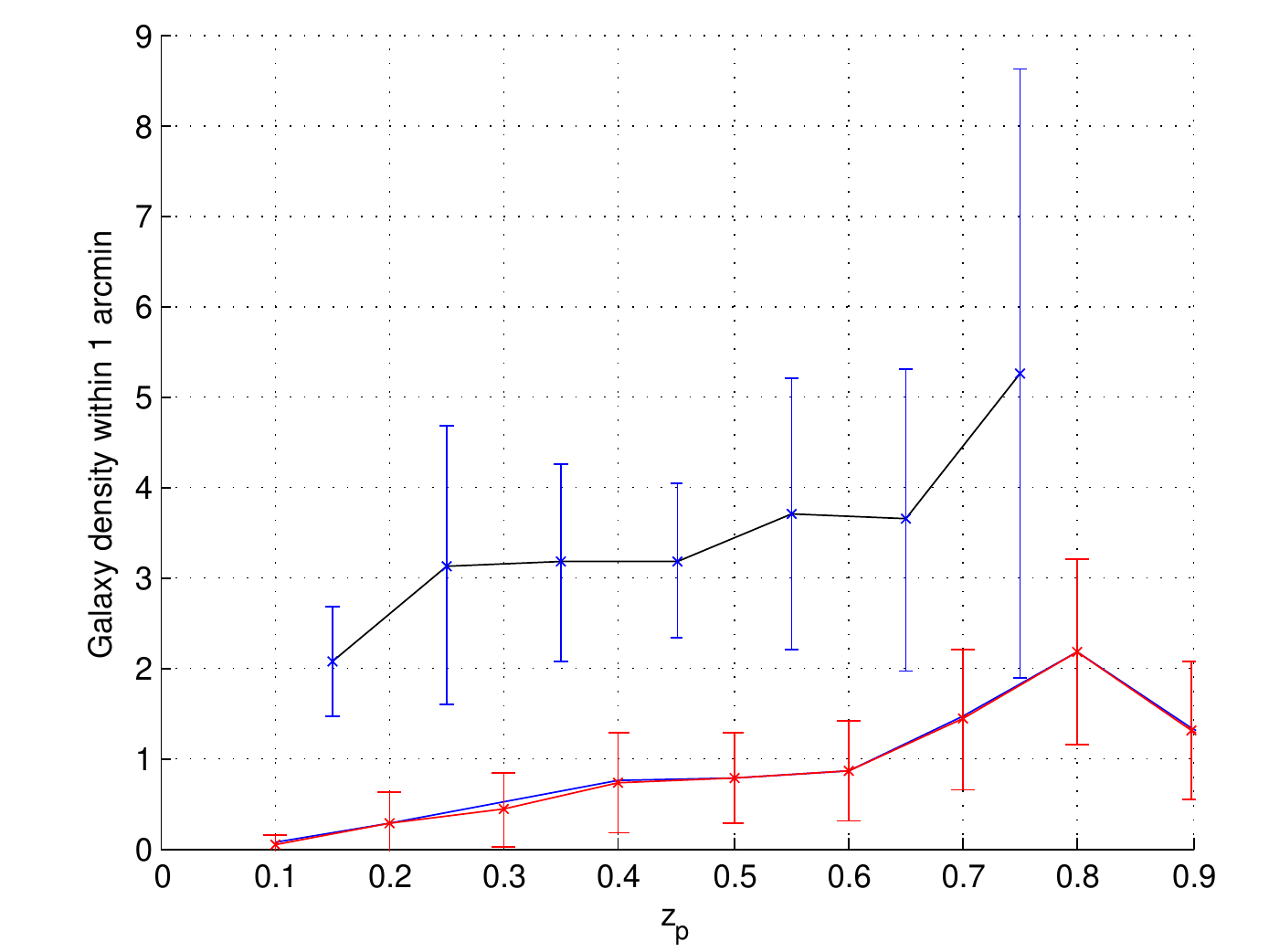}
\caption{The upper curves represent the mean galaxy density of cluster galaxies within within 500 kpc (left) and 1 arcmin (right) for 45 known clusters. The lower curves in both figures indicate the mean density for galaxies in 100 random positions before and after 3$\sigma$ clipping, respectively. For all curves the standard deviation values are indicated by the error bars.}
\label{Fig:DensityBoth}
\end{figure*}

The cluster sample includes 45 clusters with spectroscopic redshifts in the range of 0.1-0.8. We divided this range into seven redshift bins of 0.1 width. Then we computed $\mu_{500kpc}$, $\sigma_{500kpc}$, $\mu_{1arcmin}$ and $\sigma_{1arcmin}$ for those clusters in each redshift bin. Table~\ref{Table:DensityKnown} lists the redshift bins for the cluster sample, number of clusters in each bin and their mean galaxy density statistics. ~Fig.~\ref{Fig:DensityBoth} shows the mean galaxy densities (upper curve) of the 45 clusters distributed over the 7 redshift bins. The standard deviations are also indicated in the figure as error bars.        

For each random location, we create similar redshift bins as the ones used in the cluster sample. We compute the scale at these redshift bins and determine a radius of 500 kpc as well as 1 arcmin. Then, we select galaxies according to same criteria used in the cluster sample (sec. 5.2), i.e. in the same redshift interval, redshift and magnitude cuts, and within 500 kpc or 1 arcmin from the random position. We calculate the mean and standard deviation for the galaxy density in 100 positions at each redshift. We notice that, in some positions, there are over-densities of galaxies due to presence of a cluster or filament related to a cluster.  Therefore, we do 3$\sigma$ clipping to avoid these locations.

Fig.~\ref{Fig:DensityBoth} shows the mean galaxy densities (with and without 3$\sigma$ clipping) as a function of redshift for the random positions (lower curves) and their standard deviation values (error bars). As can be seen in the figure, the mean galaxy densities for the known clusters are higher than those of random positions for all redshift ranges, which indicates the ability of our algorithm to find true clusters. 

\addtocounter{table}{+1}
\begin{table*}[t]
\caption{Mean galaxy densities for known clusters in literature (45 systems). $\mu_{500kpc}$, $\sigma_{500kpc}$, $\mu_{1arcmin}$ and $\sigma_{1arcmin}$ are the mean and standard deviations of galaxy densities within 500 kpc and 1 arcmin, respectively. Cluster counts are the number of clusters within the corresponding redshift ranges.
\label{Table:DensityKnown}}
\centering
\begin{tabular}{ccccccc}
\tableline
\multicolumn{1}{c}{Bin} &
\multicolumn{1}{c}{Redshift} &
  \multicolumn{1}{c}{$\mu_{500kpc}$} &
  \multicolumn{1}{c}{$\sigma_{500kpc}$} &
  \multicolumn{1}{c}{$\mu_{1arcmin}$} &
  \multicolumn{1}{c}{$\sigma_{1arcmin}$} &
    \multicolumn{1}{c}{Clusters}\\
    \multicolumn{1}{c}{ID} &
\multicolumn{1}{c}{range} &
  \multicolumn{1}{c}{($arcmin^{-2}$)}&
  \multicolumn{1}{c}{} &
\multicolumn{1}{c}{($arcmin^{-2}$)}&
  \multicolumn{1}{c}{} &
\multicolumn{1}{c}{counts }\\

  \tableline
  \tableline
  1 & $0.1-0.2$ & $0.72$ & 0.33&   2.07 &0.61& 4\\
  2 & $0.2-0.3$ & $1.82$ &0.83 & 3.12&0.61& 10\\
  3 & $0.3-0.4$ & $1.97$ &0.62 &3.16  &1.09& 12\\
  4 & $0.4-0.5$ & $2.63$ & 0.71 &  3.18&0.86& 7\\
  5 & $0.5-0.6$ & $3.19$ & 1.47  & 3.71&1.50& 3\\
  6 & $0.6-0.7$ & $3.30$ & 1.50 & 3.64  &1.67& 7\\
  7 & $0.7-0.8$ & $4.39$ &  2.91& 5.25 &3.38&2\\
  \tableline
   total &  &  & & & &45\\
  \tableline
  \end{tabular}
  \end{table*}

\subsection{Investigating non optically confirmed clusters}
\label{SubSec:ResNonConfirmed}

We apply the clustering algorithm on 40 X-ray selected cluster candidates that have neither confirmed optical counterparts nor redshift estimates in the literature. For each candidate, our cluster finding algorithm outputs $\hat{z_p}$, $\hat{z_s}$ (when available) and cluster members within 500 kpc and 1 arcmin from the X-ray centers. Then, the galaxy densities at the cluster photometric redshift are computed and compared to the values in Table~\ref{Table:DensityKnown}. For the same redshift, if the galaxy density within 1 arcmin of a candidate exceeds ($\mu_{1arcmin}-\sigma_{1arcmin}$) and also the galaxy density within 500 kpc exceeds ($\mu_{500kpc}-\sigma_{500kpc}$), we consider the candidate detection as a cluster. If only the galaxy density within 1 arcmin excceds the referenced value ($\mu_{1arcmin}-\sigma_{1arcmin}$), we check for spectroscopic redshifts and if available we confirm the existence of a cluster. 

Among 40 candidate clusters, 11 clusters are confirmed by our algorithm based on the previously mentioned conditions. There is one distant cluster candidate (3XMM J001738.1-005150) that is along the line of sight of another nearby system and the algorithm detected the nearby one. Therefore, we do not include it in the results. This yields 10 identified clusters. 

We have noticed that 2 candidates of the remaining systems did not pass any of our galaxy density selection conditions but they have spectroscopic confirmation for one cluster galaxy. By inspecting these systems through the navigating tool at SDSS we found that the cluster galaxy with spectroscopic redshift is coinciding with the X-ray center. Therefore, we believe that they are galaxy groups and their redshift estimates from our algorithm are right. These systems are also added to our results with a note about confirmation through visual inspection which are 3XMM J232853.8+000540 (at $z_s$=0.4428) and 3XMM J022931.9+004459 (at $z_s$=0.61858). 

\begin{figure}[t]
\centering
\includegraphics[scale=0.38]{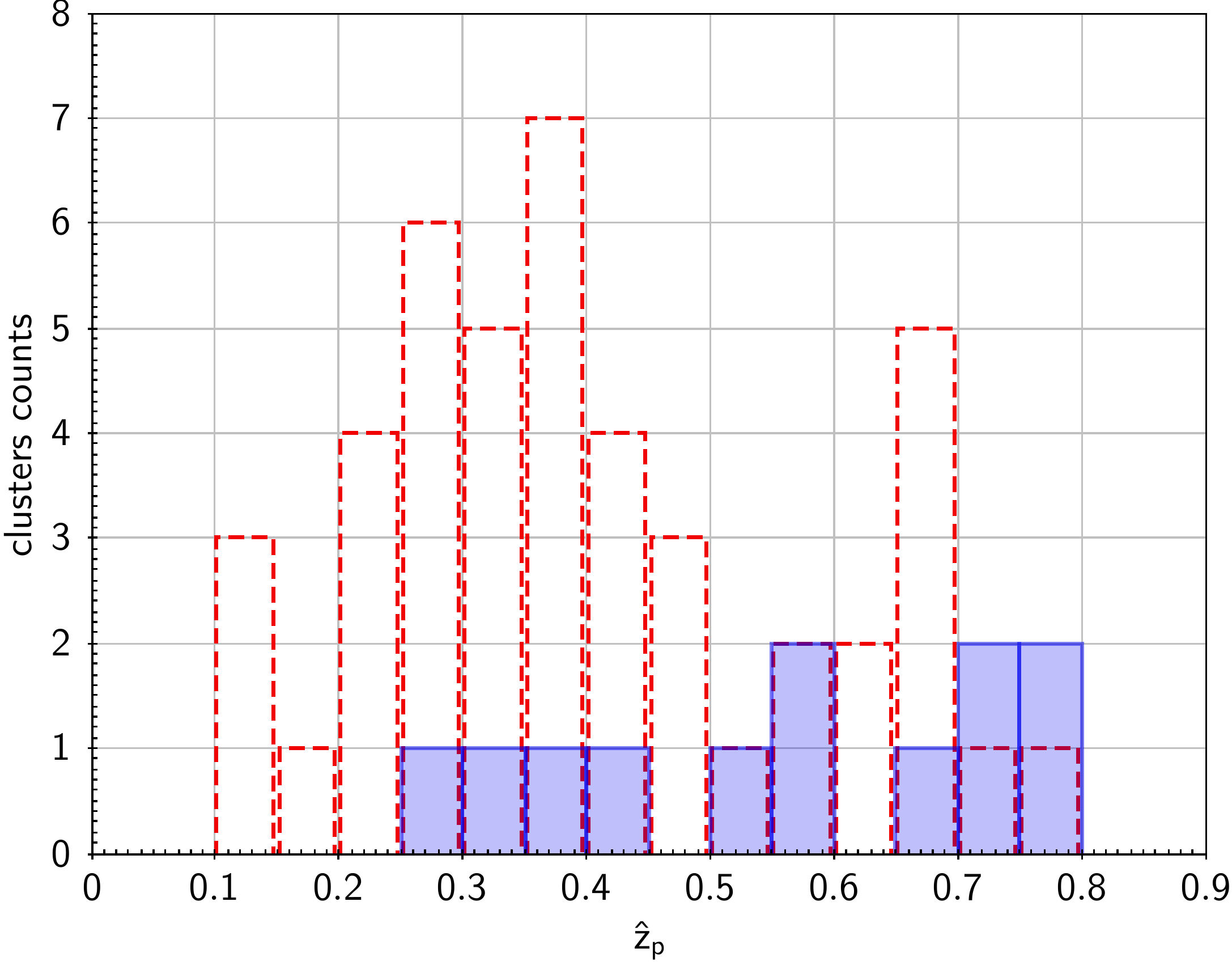}
\caption{Histogram of clusters photometric redshifts ($\hat{z_p}$) for the cluster sample (45 clusters) in red dotted bars and the new identified clusters (12 clusters) in blue bars.}
\label{Fig:GalaxyDistZp}
\end{figure}

Fig.\ref{Fig:GalaxyDistZp} shows the photometric redshift distribution of the cluster sample (45 objects) in dotted red bars with a median of 0.36. The blue bars shows the same distribution but for the 12 newly identified clusters by our algorithm with a median of 0.57.

Table~\ref{Table:Test} lists the properties of the newly confirmed clusters (12 clusters) after applying our algorithm to the cluster candidate sample (40 clusters). The columns are the same as in Table~\ref{Table:Sample}. These new 12 clusters are in a photometric redshift range from 0.29 to 0.76 with a median of 0.57. Among those 12 clusters, 7 have spectroscopic redshifts.

As we expected, the high redshift clusters at $z>0.8$ are still unconfirmed due to the magnitude cut used in our algorithm. For those high redshift clusters, we need to use deep NIR data together with current magnitudes from S82 data.

\section{Conclusion and Future Work}
\label{Sec:Dis}
A galaxy cluster finding algorithm based on spectral clustering technique is developed ({\url{https://github.com/1680/Journal-manuscript-code.git}}). The spectral clustering algorithm searches for clusters using the magnitude and color features. We apply the algorithm on a sample of 45 clusters with spectroscopic redshifts in the range of 0.1-0.8. These clusters are identified in the framework of the 3XMM/SDSS Stripe 82 galaxy cluster survey \citep{Takey2016}. Our cluster finding algorithm identified the optical counterparts of these systems and reported their photometric redshifts, which agree well with the published ones. We also apply our algorithm on a sample of 40 X-ray cluster candidates from the same cluster survey and optically confirmed 12 systems with photometric redshift range from 0.29 to 0.76 and a median of 0.57. The remaining candidates are expected to be at high redshift and therefore we need more data than the ones used in our analysis to confirm them.  

The spectral clustering algorithm is easy to implement and solve which makes it a good choice for solving different problems. Also, this algorithm does not cluster the observations (galaxies) itself as the $K$-means algorithm does, but works on the similarity matrix created from the relations between different galaxies in a specific space. This results in the ability of the algorithm to identify galaxy clusters by the connectivity of their members as well as their densities \citep{Lux2007, Kannan2004}.

The parameters inside the spectral clustering algorithm can be adaptive to self tune the algorithm such as $\sigma$, the local scaling parameter and $k$, the number of output clusters \citep{Manor2004}. The algorithm also can be a learning one as it can learn the similarity matrix from the data \citep{Bach2004}. As a first application of the spectral clustering to find galaxy clusters, we use the basic algorithm according to \citep{Ng2002}. The self-tuning and learning versions cab be used in future work.


 \acknowledgments
 
The Ministry of Higher Education (MoHE) of Egypt supported this research through a Ph.D. fellowship. I expand my deepest gratitude to Prof. Ahmed El-Mahdy, CSE Dept., EJUST University, and Prof. Gamal Bakr Ali, Astronomy Dept., NRIAG, for their continuous support and help.

Ali Takey acknowledges the postdoctoral fellowship supported by the Egyptian Science and
Technology Development Fund (STDF) and the French Institute in Egypt (IFE).

Funding for SDSS-III has been provided by the Alfred P. Sloan 
Foundation, the Participating Institutions, the National Science Foundation,
and the U.S. Department of Energy. The SDSS-III web site is 
\url{http://www.sdss3.org/}. 
SDSS-III is managed by the Astrophysical Research 
Consortium for the Participating Institutions of the SDSS-III Collaboration 
including the University of Arizona, the Brazilian Participation Group, 
Brookhaven National Laboratory, University of Cambridge, University of 
Florida, the French Participation Group, the German Participation Group, 
the Instituto de Astrofisica de Canarias, the Michigan State/Notre 
Dame/JINA Participation Group, Johns Hopkins University, Lawrence Berkeley 
National Laboratory, Max Planck Institute for Astrophysics, New Mexico State 
University, New York University, Ohio State University, Pennsylvania State 
University, University of Portsmouth, Princeton University, the Spanish 
Participation Group, University of Tokyo, University of Utah, Vanderbilt 
University, University of Virginia, University of Washington, and Yale University.






\bibliographystyle{aa}
\bibliography{refbib_thesis}




\begin{appendix}
\label{Append:A}
\begin{center}
{\bf Appendix (A): Flowchart of the galaxy cluster finding algorithm~\ref{Alg:MyAlg}}
\newline
\newline
\newline
\begin{tikzpicture}[auto]
\node (start) [startstop] {Start};
\node (pro0) [process, below of=start] {For the proposed X-ray candidate: select constrained galaxies, prepare feature vector for each galaxy, generate similarity matrix (steps 1.1 ... 1.3 in the algorithm)};
\draw [arrow] (start) -- (pro0);
\node (in1) [io, below of=pro0] {Galaxies similarity matrix};
\draw [arrow] (pro0) -- (in1);
\node (pro1) [process, below of=in1] {Spectral clustering (step 1.4 in the algorithm)};
\draw [arrow] (in1) -- (pro1);
\node (out1) [io, below of=pro1] {Target cluster and fore-,back-ground cluster};
\draw [arrow] (pro1) -- (out1);
\node (pro2) [process, below of=out1] {Calculate the tentative redshift from the target cluster galaxies photometric redshifts (steps 1.5 and 1.6 in the algorithm)};
\draw [arrow] (out1) -- (pro2);
\node (pro3) [process, below of=pro2] {Fine tune the tentative redshift (step 1.7 in the algorithm)};
\draw [arrow] (pro2) -- (pro3);
\node (out2) [io, below of=pro3] {$\hat{z_p}$, $\hat{z_s}$ and members of the galaxy cluster};
\draw [arrow] (pro3) -- (out2);
\node (dec1) [decision, right of=out2] {More X-ray Candidates?};
\draw [arrow] (out2) -- (dec1);
\node (stop) [startstop, below of=out2] {Stop};
\draw [arrow] (dec1) |- node [anchor=north] {No} (stop);
\draw [arrow] (dec1) |- node [anchor=south] {Yes} (pro0);
\label{Chart:MyChart}
\end{tikzpicture}
 \end{center}
\end{appendix}



\addtocounter{table}{-2}
\begin{landscape}
\begin{table*}
\centering
\footnotesize
\caption{Results from our cluster finding algorithm for the cluster sample (45 objects) with redshift measurements in literature.\label{Table:Sample}}
\begin{tabular}{@{}crrrrrrrrrrr@{}}
\tableline
\tableline
\multicolumn{1}{c}{IAUNAME} &
  \multicolumn{1}{c}{$\alpha_{Xray}$} &
  \multicolumn{1}{c}{$\delta_{Xray}$} &
\multicolumn{1}{c}{DETID} &
  \multicolumn{1}{c}{$\alpha_{BCG}$} &
  \multicolumn{1}{c}{$\delta_{BCG}$} &
  \multicolumn{1}{c}{$n_{z_p}$} &
  \multicolumn{1}{c}{$\hat{z_p}$} &
  \multicolumn{1}{c}{$n_{z_s}$} &
  \multicolumn{1}{c}{$\hat{z_s}$} &
  \multicolumn{1}{c}{$z_{pub}$} \\

\tableline
\tableline
    3XMM J001115.5+005152 & 2.8147 & 0.86462 & 104037603010094 & 2.813276 & 0.865525 & 16  & 0.368981 & 1  & 0.364701 & 0.3622\\
  3XMM J001737.3-005240 & 4.40608 & -0.87794 & 104037601010003 & 4.406702 & -0.878333 & 38 & 0.219055 & 5 & 0.212995 & 0.2141\\
  3XMM J002223.3+001201 & 5.59736 & 0.20036 & 104070301010041 & 5.610213 & 0.193065 & 14 & 0.576964 & 0 & - & 0.2789\\
  3XMM J002314.4+001200 & 5.81017 & 0.20016 & 104070301010056 & 5.81153 & 0.199448 & 18 & 0.26206 & 3 & 0.25799 & 0.2597\\
  3XMM J003838.0+004351 & 9.65851 & 0.73108 & 102036901010028 & 9.659185 & 0.731283 & 27 & 0.719093 & 0 & - & 0.6955\\
  3XMM J003840.3+004747 & 9.66813 & 0.79659 & 102036901010085 & 9.664727 & 0.794459 & 26 & 0.678965 & 0 & - & 0.5553\\
  3XMM J003922.4+004809 & 9.84359 & 0.80277 & 102036901010023 & 9.846039 & 0.792243 & 30 & 0.4019 & 3 & 0.415292 & 0.4145\\
  3XMM J003942.2+004533 & 9.92597 & 0.75926 & 102036901010017 & 9.931303 & 0.764728 & 14 & 0.364147 & 4 & 0.289587 & 0.4156\\
  3XMM J004231.0+005112 & 10.6293 & 0.85336 & 100900702010087 & 10.630933 & 0.850208 & 19 & 0.139971 & 5 & 0.155954 & 0.1579\\
  3XMM J004252.5+004300 & 10.71892 & 0.71692 & 100900702010056 & 10.722837 & 0.712743 & 19 & 0.338188 & 3 & 0.27034 & 0.2697\\
  3XMM J004334.1+010107 & 10.89187 & 1.01811 & 100900702010050 & 10.896067 & 1.019719 & 26 & 0.191657 & 3 & 0.155714 & 0.2\\
  3XMM J004350.6+004731 & 10.96114 & 0.79216 & 100900702010052 & 10.957893 & 0.790219 & 24 & 0.589527 & 0 & - & 0.4754\\
  3XMM J004401.4+000644 & 11.00583 & 0.11226 & 103035622010028 & 11.005035 & 0.114357 & 29 & 0.216996 & 3 & 0.218679 & 0.2187\\
  3XMM J010606.7+004925 & 16.52804 & 0.82388 & 101508702010016 & 16.529263 & 0.819496 & 39 & 0.267828 & 4 & 0.254256 & 0.2564\\
  3XMM J010610.0+005108 & 16.54172 & 0.85226 & 101508702010012 & 16.543246 & 0.855698 & 41 & 0.281673 & 6 & 0.261641 & 0.2566\\
  3XMM J015917.1+003010 & 29.82144 & 0.503 & 101016402010005 & 29.823164 & 0.518687 & 28 & 0.386031 & 1 & 0.383999 & 0.382\\
  3XMM J020019.2+001932 & 30.08002 & 0.32564 & 101016402010018 & 30.081002 & 0.324922 & 8 & 0.627479 & 1 & 0.682471 & 0.6825\\
  3XMM J021012.6-001439 & 32.55263 & -0.24443 & 102007306010070 & 32.551035 & -0.247333 & 18 & 0.301448 & 1 & 0.282752 & 0.2828\\
  3XMM J021045.8-002156 & 32.69099 & -0.36574 & 102007306010018 & 32.690059 & -0.366493 & 18 & 0.36492 & 1 & 0.314387 & 0.31\\
  3XMM J022825.8+003203 & 37.1078 & 0.53441 & 106524006010012 & 37.107716 & 0.53406 & 21 & 0.415019 & 1 & 0.395171 & 0.3952\\
  3XMM J022830.5+003032 & 37.12738 & 0.50907 & 106524006010017 & 37.126887 & 0.509941 & 11 & 0.691346 & 0 & - & 0.72\\
  3XMM J023058.5+004327 & 37.74395 & 0.72431 & 106524008010044 & 37.730579 & 0.73099 & 15 & 0.479382 & 3 & 0.472347 & 0.4727\\
  3XMM J025846.5+001219 & 44.69388 & 0.20555 & 106064313010011 & 44.694118 & 0.205125 & 17 & 0.259239 & 4 & 0.252081 & 0.2589\\
  3XMM J025932.5+001353 & 44.88574 & 0.23161 & 106064313010004 & 44.850586 & 0.251131 & 32 & 0.206248 & 1 & 0.194413 & 0.192\\
  3XMM J030145.7+000323 & 45.44072 & 0.05659 & 100411701010097 & 45.436488 & 0.05406 & 19 & 0.678286 & 0 & - & 0.69\\
  3XMM J030205.6-000001 & 45.52347 & -5.3E-4 & 100411701010033 & 45.522255 & -0.013152 & 9 & 0.617979 & 0 & - & 0.65\\
  3XMM J030212.1+001107 & 45.55082 & 0.18536 & 100411701010113 & 45.548192 & 0.187502 & 13 & 0.653403 & 1 & 0.652276 & 0.6523\\
  3XMM J030317.5+001245 & 45.82296 & 0.21272 & 100411701010112 & 45.822218 & 0.217699 & 17 & 0.675781 & 0 & - & 0.59\\
  3XMM J030614.1-000540 & 46.55923 & -0.09474 & 101426101010024 & 46.558861 & -0.094392 & 9 & 0.391685 & 1 & 0.424876 & 0.4249\\
  3XMM J030617.3-000836 & 46.57206 & -0.14361 & 101426101010022 & 46.546886 & -0.138734 & 33 & 0.110217 & 24 & 0.124163 & 0.1093\\
  3XMM J030633.1-000350 & 46.63804 & -0.06408 & 101426101010059 & 46.685936 & -0.024054 & 28 & 0.112077 & 22 & 0.129189 & 0.1235\\
  3XMM J030637.3-001801 & 46.6557 & -0.30054 & 102011201010042 & 46.652955 & -0.302152 & 17 & 0.452141 & 1 & 0.457565 & 0.4576\\
  3XMM J033446.2+001710 & 53.69279 & 0.28618 & 104023202010027 & 53.693565 & 0.285401 & 17& 0.311778 & 1 & 0.326141 & 0.3261\\
  3XMM J035416.9-001003 & 58.5706 & -0.16751 & 101349209010028 & 58.583058 & -0.187765 & 21 & 0.273699 & 0 & - & 0.21\\
  3XMM J221211.0-000833 & 333.04618 & -0.14275 & 106553468400021 & 333.047742 & -0.139106 & 20 & 0.361821 & 1 & 0.364674 & 0.3643\\
  3XMM J221422.1+004712 & 333.59226 & 0.7868 & 106553468390009 & 333.593067 & 0.784941 & 21 & 0.342137 & 2 & 0.320206 & 0.3202\\
  3XMM J221449.2+004707 & 333.70526 & 0.78535 & 106553468390023 & 333.706222 & 0.785063 & 21 & 0.321656 & 2 & 0.317685 & 0.3171\\
  3XMM J222144.0-005306 & 335.43347 & -0.88513 & 106700202010013 & 335.432211 & -0.884142 & 11 & 0.361478 & 2 & 0.335458 & 0.3363\\
  3XMM J232540.3+001447 & 351.41827 & 0.2466 & 106524009010093 & 351.420944 & 0.248603 & 9 & 0.804074 & 0 & - & 0.79\\
  3XMM J232613.8+000706 & 351.55771 & 0.11844 & 106524009010085 & 351.556101 & 0.114583 & 16 & 0.470468 & 1 & 0.458356 & 0.3844\\
  3XMM J232742.1+001406 & 351.9255 & 0.23522 & 106524010010078 & 351.906263 & 0.234037 & 11 & 0.445455 & 1 & 0.445131 & 0.4441\\
  3XMM J232809.0+001116 & 352.03771 & 0.18778 & 106524010010043 & 352.038511 & 0.185933 & 9 & 0.261988 & 1 & 0.276827 & 0.278\\
  3XMM J232925.6+000554 & 352.35668 & 0.09849 & 106524011010056 & 352.355668 & 0.099429 & 17 & 0.400789 & 1 & 0.401929 & 0.4021\\
  3XMM J233138.1+000738 & 352.90912 & 0.12725 & 106524014010034 & 352.908655 & 0.128404 & 20 & 0.243618 & 1 & 0.223817 & 0.2344\\
  3XMM J233328.1-000123 & 353.36739 & -0.02308 & 106524013010039 & 353.366237 & -0.022758 & 11 & 0.529217 & 1 & 0.512008 & 0.462\\
\tableline
\end{tabular}
\end{table*}
\end{landscape}

\addtocounter{table}{+1}
\begin{landscape}
\begin{table*}
\normalsize	
\centering
\caption{Results from our cluster finding algorithm for the  X-ray cluster candidates sample.\label{Table:Test}}
\begin{tabular}{@{}crrrrrrrrrrr@{}}
\tableline
\tableline
\multicolumn{1}{c}{IAUNAME} &
  \multicolumn{1}{c}{$\alpha_{Xray}$} &
  \multicolumn{1}{c}{$\delta_{Xray}$} &
\multicolumn{1}{c}{DETID} &
  \multicolumn{1}{c}{$\alpha_{BCG}$} &
  \multicolumn{1}{c}{$\delta_{BCG}$} &
  \multicolumn{1}{c}{$n_{z_p}$} &
  \multicolumn{1}{c}{$\hat{z_p}$} &
  \multicolumn{1}{c}{$n_{z_s}$} &
  \multicolumn{1}{c}{$\hat{z_s}$} &\\
\tableline
\tableline
   3XMM J000436.8+000148 & 1.15346 & 0.03028 & 103057510010070 & 1.1426 & 0.023643 & 12 & 0.697108 & 0 &- \\
  3XMM J001200.5+005233 & 3.00223 & 0.87606 & 104037603010026 & 3.0117 & 0.87215 & 10 & 0.764716 & 0 &- \\
  3XMM J002905.7-001639 & 7.27381 & -0.27767 & 104031601010027 & 7.2567 & -0.28537 & 9 & 0.76493 & 0 &- \\
  3XMM J021021.7-000721 & 32.59055 & -0.12264 & 102007306010046 & 32.581 & -0.1172 & 7 & 0.290746 & 1 & 0.27801\\
  3XMM J022740.0+003926 & 36.91706 & 0.65747 & 106524006010067 & 36.915 & 0.66742 & 8 & 0.568811 & 1 & 0.5612\\
  3XMM J022834.3+004447 & 37.1433 & 0.74639 & 106524006010070 & 37.154 & 0.73578 & 7 & 0.711876 & 0 &- \\
  3XMM J022931.9+004459 \tablenotemark{*}& 37.38322 & 0.7499 & 106524007010092 & 37.385 & 0.75554 & 5 & 0.579169 & 1 & 0.61858\\
  3XMM J030652.9-001121 & 46.7205 & -0.18939 & 102011201010049 & 46.73 & -0.19145 & 14 & 0.37354 & 1 & 0.36548\\
  3XMM J030653.7-000309 & 46.72399 & -0.05272 & 101426101010088 & 46.718 & -0.048222 & 8 & 0.746635 & 0 & -\\
  3XMM J222116.2-010031 & 335.31774 & -1.00882 & 106700202010014 & 335.32 & -1.0125 & 15 & 0.338128 & 1 & 0.31634\\
  3XMM J232850.3+001356 & 352.20977 & 0.23222 & 106524011010077 & 352.21 & 0.23759 & 10 & 0.514666 & 1 & 0.44209\\
  3XMM J232853.8+000540 \tablenotemark{*} & 352.22439 & 0.09466 & 106524011010058 & 352.22 & 0.093566 & 8 & 0.424996 & 1 & 0.4428\\

\tableline
\end{tabular}
\tablenotetext{*}{The results of our algorithm for this cluster is confirmed through visual inspection}
\end{table*}
\end{landscape}




\end{document}